# Fractal Location and Anomalous Diffusion Dynamics for Oil Wells from the KY Geological Survey

Keith Andrew
Department of Physics and Astronomy
Keith.Andrew@wku.edu

Karla M. Andrew
Center for Water Resources Studies
Karla.Andrew@wku.edu

Kevin A. Andrew
Carol Martin Gatton Academy of Math and Science
Kevin.Andrew@wku.edu

Western Kentucky University
Bowling Green, KY 42101

Abstract

Utilizing data available from the Kentucky Geonet (KYGeonet.ky.gov) the fossil fuel mining locations created by the Kentucky Geological Survey geo-locating oil and gas wells are mapped using ESRI ArcGIS in Kentucky single plain 1602 ft projection. This data was then exported into a spreadsheet showing latitude and longitude for each point to be used for modeling at different scales to determine the fractal dimension of the set. Following the porosity and diffusivity studies of Tarafdar and Roy[1] we extract fractal dimensions of the fossil fuel mining locations and search for evidence of scaling laws for the set of deposits. The Levy index is used to determine a match to a statistical mechanically motivated generalized probability function for the wells. This probability distribution corresponds to a solution of a dynamical anomalous diffusion equation of fractional order that describes the Levy paths which can be solved in the diffusion limit by the Fox H function ansatz.

## I. Introduction

Many observed patterns in nature exhibit a fractal like pattern at some length scale, including a variety of mineral deposits[2,3]. Here we analyze GIS data created by the KY Geological Survey and available through KY Geonet in terms of an introductory statistical model describing the locations of oil and gas deposits as mapped by their respective wells. The model indicates there can exist a power law critical exponent and that the probability distribution has a stable Levy index. Different values of these indices correspond to different probability distributions and underlying dynamics. The index provides a link to the fractal structure of the underlying point distribution of the wells and in principle provides insight into the dynamics of mineral structure formation when viewed as formed by diffusion of a multicomponent thermoviscous fluid. A particularly useful model for this type of analysis is the diffusion limited aggregation, DLA, model and its variants which has been used to understand[2] colloidal aggregates, bacterial colonies, corals, percolation[4], and in the case of porous media, the formation of sedimentary rocks [5,6]. Such a model gives insight into water flow and sedimentary rock deposition as developed by Tarafdar and Roy[7] and into an anomalous diffusion and percolation model of Chen.[8]

Here we use the known deposits from well locations to determine the fractal dimension of the deposit structure and to model the probability distribution of these wells. In the next section we describe the data set used from the Kentucky Geological Survey, the third section describes a generalization of the Levy stable probability function derived by Mekjian[9] that is useful for optimizing the function of interest, in the fourth section the fractal dimension, Levy index and dynamics are described for the data set and the Fox H function ansatz is used in the diffusion limit to arrive at expressions that are valid in the anomalous regime, and the conclusions are in the final section.

## II. Data from the Kentucky Geological Survey

Data was downloaded from the Kentucky Geological Survey at the University of Kentucky (http://www.uky.edu/KGS/gis/geology.htm). The data includes nearly 140,000 oil and gas wells within the state of Kentucky. Some duplication has occurred due to modification of wells and data maintenance.

Files are projected using the North American Datum of 1983. Location data in the attributes supplied in both shapefiles include latitude and longitude in both North American data of 1927 and 1983 (NAD27 and NAD83). For the use of this study the NAD83 latitude and longitude coordinates were used. Further study could be done with this data set using the Org Result field showing the original location and features of the well. Files were mapped using ESRI ArcGIS and then records were exported to include the NAD83 latitude and longitude to a comma delimited file. This file was then imported to an Access database. A map of the wells with a statewide partition is shown in Fig. 1 and a box count distribution is shown on the right.

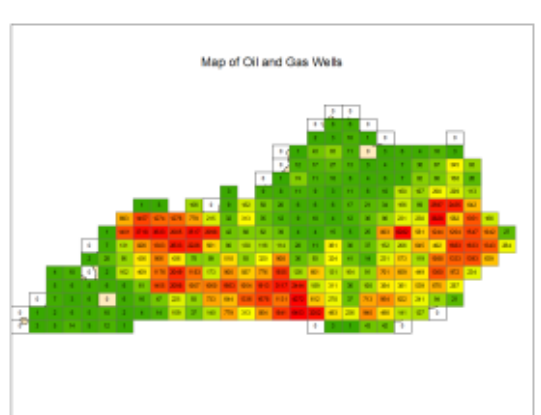

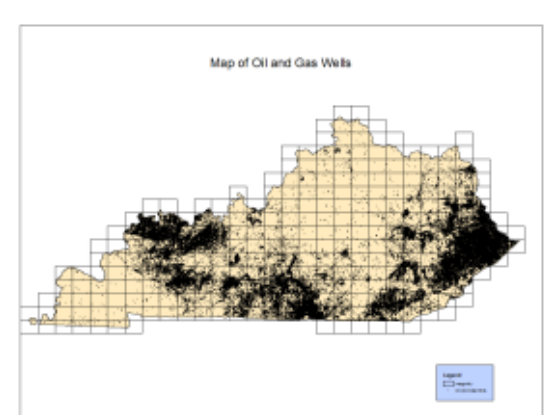

Fig. 1 Map of the grid and oil and gas wells in Kentucky on left and the box count distribution on the right.

To analyze the data the records of the mines were grouped into bins based on latitude and longitude, dividing the map into square intervals Δx wide and Δy high. This was accomplished using SQL queries within Access. The file allows any interval size to be selected. The resulting number tells the average number of records within bins based on the chosen interval. Counting was done for a variety of scales in box size to determine the probability functions.

### III. General Levy Stable Probability Functions

The general point probability function applied to wells, the GPF, is defined as the probability of finding a well inside a sphere or radius R, randomly placed within a sample size greater than R. For spherical volume elements the GPF can be expressed as:

$$P_o(R) = \exp\left[ -\sum_{p=1}^{\infty} \frac{\left(-\overline{N}(R)\right)^p}{p!} \overline{\xi}_p \right] \tag{1}$$

where R is the sphere radius, $\overline{N}$ is the average number of wells in the sphere, and $\overline{\xi}$ is the volume averaged p-point correlation function defined as

$$\overline{\xi}_p = \frac{\int \xi_p dV}{\int dV} . \tag{2}$$

Following Meikjain[8] we will apply the hierarchical ansatz to relate all higher order correlation functions to the two point function, ξ, by

$$\xi_p = S_p \xi^{p-1}, \quad p \geq 3 \tag{3}$$

for scaling coefficients $S_p$ giving a GPF as:

$$P_o = \exp\left[ -\sum_{p=1}^{\infty} \frac{\left(-\overline{N}\right)^p}{p!} S_p \overline{\xi}^{\,p-1} \right]. \tag{4}$$

Hierarchical clumping would be characterized by the formation of small clumps that are later absorbed by large scale clumps as new small clumps form. To isolate the effects of the scaling coefficients Fry[10,11] introduced the reduced generalized probability distribution, $\chi$, RGPF, with an independent variable $x = \overline{N}\overline{\xi}$ :

$$\chi(x) = -\ln(P_o)/\overline{N} = \sum_{p=1}^{\infty} \frac{S_p}{p!} (-x)^{p-1} . \tag{5}$$

As shown by Mekjian[9] some simple analytical void distributions can be effectively understood by choosing a model that allows for the calculation of the scaling coefficients, $S_p$. Several of these are collected in the table below and have been used extensively in other models[11]:

| Model | $S_p$ | GPF | RGPF |
|---|---|---|---|
| Poisson | p=1, $S_1$=1, all others zero | $P_o = e^{-\bar{N}}$ | $\chi = 1$ |
| Gaussian | Only p=2 nonzero | $P_o = e^{-\bar{N}^2\bar{\xi}/2}$ | $\chi = 1 - x/2$ |
| Minimal | Sp=1 for all p | $P_o = e^{-\bar{N}(1+e^{-x})/2}$ | $\chi = (1+e^{-x})/2$ |
| Thermodynamic | Sp=(2p-3)!! | $P_o = e^{-\left[(1+2x^{1/2}-1)\right]/x}$ | $\chi = \left[(1+2x)^{1/2}-1\right]/(x)$ |
| Negative Binomial | Sp=(p-1)! | $P_o = \left(\frac{1}{1+x}\right)^{1/\bar{\xi}}$ | $\chi = \ln(1+x)/(x)$ |

Table 1. Some Analytic Generalized Probability Functions

These results may be expressed in terms of a generalized generating functional, which explicitly utilizes a parameter "a" that can be related to the Levy stability index α when the probability distribution is expressed as a hypergeometric function[8]. Here the relationship is simply that a = 1-α. This formulation leads to a convenient power law expression for the probability distribution at large x given by:

$$P \sim x^{-(1+\alpha)} . \tag{7}$$

For $\alpha < 2$ the probability function will converge to a Levy stable stochastic process of index α. Only two special cases are known to be expressible in closed form; α =1 yields a Cauchy distribution and α = 2 is a Gaussian distribution. For α contained in the interval (0,2] α is the Levy index of stability or the tail index, for these cases if the autocorrelation remains near zero the stochastic variables are hierarchical and obey a simple scaling law[9]. Here we will use Eq.(7) to fit our data on empty regions for a generalized Zipf's Law. Our analysis will help determine the value of the Levy index in the stable region of α. The RGPF distribution function and scale factors for any hierarchical model are then

$$\begin{aligned} P_o &= e^{-\bar{N}\chi_a} \\ \chi_a &= \frac{1}{(1-a)(\xi_2\bar{N}/a)}\left[\left(1+\frac{\xi_2\bar{N}}{a}\right)^{1-a}-1\right] \\ \xi_p &= S_p\xi_2^{p-1} \qquad S_p = \frac{\Gamma(a+p-1)}{\Gamma(a)a^{p-1}} \end{aligned} \quad . \tag{8}$$

Here we will use Eq.(8) to determine the parameter value of "a" from the Geonet observational data. Notice that the scaling factors might increase or decrease with increasing p depending upon the value of a. This formulation of the distribution function has the distinct advantage of having a clear statistical mechanical derivation with connections to percolation theory and aggregate sedimentation theory[1] providing physical insight into the hierarchical structure formation models used to understand the overall distribution of oil and gas deposits. The tail of a Levy distribution characterizes the exponents associated with a scaling power law of fragmentation as it applies in powder formation, soil physics, nuclear physics, genetics, sprays, fuel combustion, brittle fracture, geologic sedimentation, etc. as a system undergoes a phase transition from uniform to fragmented, as in mineral deposit formation. When finite size scaling exists and near criticality universality exists then the Levy index characterizes the system undergoing fragmentation[8]. Here we find the value of the Levy index from observational data sets available at KY Geonet by numerically fitting the power law and RGPF.

## IV. Fractal Dimension of the Wells and Zipf's Law for No Wells: $\overline{N} = 0$

We use the box counting dimension or capacity method to determine the fractal dimension of the well dataset. Let the number of wells in a volume of radius r, be given by N(r), then, for initial count k the dimension[2] d, is given by

$$N(r) = kr^{-d}$$
$$d = d_b = \lim_{r \to 0} \frac{\log(N(r))}{\log\left(\frac{1}{r}\right)} \tag{9}$$

which may not be an integer but reproduces the integer values for appropriate data. Here we measure the fractional dimension of the geographic well distribution from the slope of the log-log plot of the number of wells at each scale, Fig. (2), using Eq.(9) and find d=1.7, this is indicative of an anomalous diffusion process.

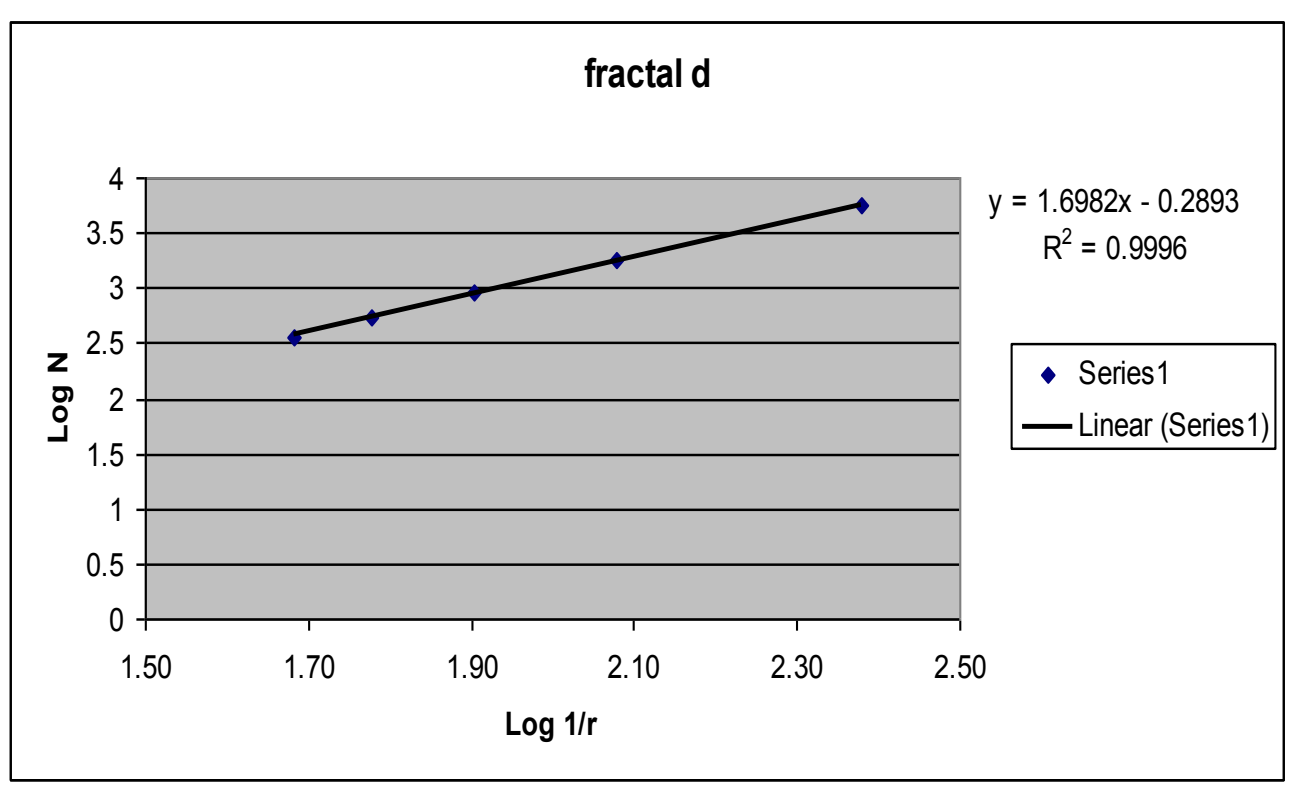

Fig. 2 Log-Log plot to determine the fractional dimension.

It is significantly simpler to calculate the distributions for the special case of no wells or voids which corresponds to $\overline{N} = 0$ in the analysis of Section III. Counting the number of areas that contain no wells for each scale we find that the empty spaces obey a generalized Zipf's Law, as in Eq.(7), of the form

$$P_n = \frac{k}{n^a} \tag{10}$$

Where the probability of finding an empty area of box size n decreases as a power law, as seen in Fig. (3).

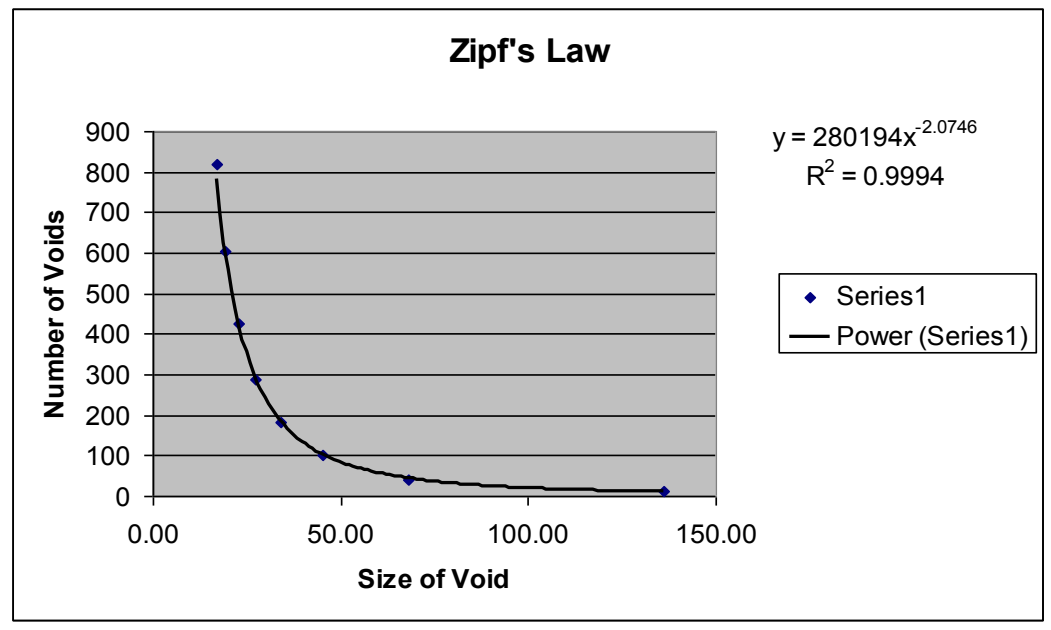

Fig. 3 Data fit to a generalized Zip's Law

This gives a Zipf power of 2.075 and a Levy index of 1.075 indicating a Levy stable probability distribution for the empty spaces containing no wells. Using the tails of the numerical fit to Eq.(9) we find a best fit for a= 2.08 or α = 1.08. Both of these methods are indicative of a value near 1.1.

Following Chen the Levy stable index can also be used to determine the order of the fractional order anomalous diffusion equation[13], that dynamically gives rise to the well structure by using the Riesz fractional derivative operator[14,15] for this system, simplified to one spatial dimension we have

$$\frac{\partial p}{\partial t} = k_\alpha \left(\frac{\partial}{\partial x}\right)^\alpha p \quad with \quad p(xt \mid x_o t_o) = \delta(x - x_o) \tag{11}$$

where solutions to this equation give values that match the sedimentary rock formation equation as a thermoviscous dissipative multicomponent fluid with diffusion coefficient $k_\alpha$ with an initial distribution given by a Dirac delta function. For integer values of α the solutions of Eq. (11) are expressible as hypergeometric function, ${}_pF_q$ and the special case of α =2 is normal diffusion. For noninteger values this equation yields a solution ansatz[16] in the diffusion limit (large |x|) that may be expressed in terms of the Fox H function as

$$p(x,t) = Na(t)H^{1,1}_{2,2}\left[ a(t)|x| \middle| \begin{matrix} (a_1, A_1) & (a_2, A_2) \\ (b_1, B_1) & (b_2, B_2) \end{matrix} \right]$$

where

$$H(z) = H^{m,n}_{p,q}\left[ z \middle| \begin{matrix} (a_1,\alpha_1),\ldots & ,(a_p,\alpha_p) \\ (b_1,\beta_1),\ldots & ,(b_p,\beta_p) \end{matrix} \right] = \frac{1}{2\pi i}\oint \frac{\prod_{j=1}^{m}\Gamma(b_j - \beta_i s)\prod_{j=1}^{n}\Gamma(1 - a_j + \alpha_j s)}{\prod_{j=n+1}^{p}\Gamma(a_j - \alpha_j s)\prod_{j=m+1}^{q}\Gamma(1 - b_j + \beta_j s)} z^s ds \tag{12}$$

with normalization constant N and scaling function $a(t) \sim Ct^{-1/\alpha}$. In this formulation a Levy stability index of 2 gives the Gaussian probability[12] distribution function. This equation is of the class used to describe anomalous diffusion where the particles can jump a large distance compared to the rms displacement and where the special case of normal diffusion is Brownian, i.e. the particles move with no memory. Here at $\alpha$=1.08 indicative of anomalous diffusion.

## V. Conclusions

From Kentucky Geonet data we have been able to determine the fractional dimension of the distribution of wells and found it to be d= 1.7 ±0.1. A corresponding analysis of the regions that contain no wells, the void regions, allows us to determine the power of a generalized Zipf's Law describing the frequency distribution of the voids. From this we ascertain the Levy stable index for the probability distribution of the voids to be $\alpha = 1.07 \pm 0.02$. This index is indicative of the order of the fractional anomalous diffusion equation which can be related to a dynamical model derivable from a thermo-viscous Navior Stokes fluid, which has a solution given by the Fox H function. By considering anomalous diffusion we have been able to find a match to the fractal structure of the wells. Our ongoing work is to model solutions to Eq.(12) at different scales to study the resulting probability distributions and compare them to the data generated from the fractal well and void structures to determine how far from normal diffusion this pattern might be. Although general solutions for stable Levy indices are expressible in terms of the Fox H function[17] they can also be expressed as a general path integral solution of a fixed noninteger order differential equation solution[18,19]. For numerical results it is easier to expand the H function as a product of infinite series as done by Srowkowski[16], we are now using theses series expansions to plot the H function results. As more Geonet data sets become available covering a wider geographical area we will extend this analysis and examine the H function stability of the data on these larger maps. If the structure is truly fractal the fractal dimension should remain the same and the pattern will be repeated at larger scales up to some cutoff value.

Acknowledgements: This work has been generously supported by a KY NASA Space Grant Consortium grant, the WKU Institute for Astrophysics and Space Science and the WKU Center for Water Resource Studies.